\documentclass[doublecol]{epl2} 
\usepackage{graphicx}% Include figure files
\usepackage{dcolumn}% Align table columns on decimal point
\usepackage{bm}% bold math
\usepackage{amssymb}
\usepackage{amsmath}
\begin{document}
\setcounter{page}{1}
\vskip 2cm
\title{On the apparent loss of predictability inside the de-Rham-Gabadadze-Tolley non-linear formulation of massive gravity: The Hawking radiation effect.}
\shorttitle{\textbf{The apparent loss of predictability in dRGT massive gravity: Hawking radiation effect}}

\author{Ivan Arraut}

\institute{Department of Physics, Osaka University, Toyonaka, Osaka 560-0043, Japan}

\abstract{I explain in a simple and compact form the origin of the apparent loss of predictability inside the dRGT non-linear formulation of massive gravity. This apparent pathology was first reported by Kodama and the author when the stability of the Schwarzschild de-Sitter (S-dS) black-hole in dRGT was analyzed. If we study the motion of a massive test particle around the S-dS solution, we find that the total energy is not conserved in the usual sense. The conserved quantity associated with time appears as a combination of the total energy and a velocity-dependent term. If the equations of motion are written in terms of this conserved quantity, then the three-dimensional motion in dRGT will not differ with respect to the same situation of Einstein gravity (GR). The differences with respect to GR will appear whenever we have a dynamical situation. I explore the Hawking radiation as an example where we can find differences between GR and dRGT.}

\maketitle

\section{Introduction}
The cosmological constant ($\Lambda$) problem, found originally inside Quantum Field theory, is just the failure to explain the observed value of the cosmological constant if it corresponds to the vacuum energy coming from the zero-point quantum fluctuations. There have been many attempts for solving the problem. In some cases by trying to find some mechanism in order to explain the no contribution of most of the modes coming from the vacuum energy \cite{1}. In other cases by modifying gravity. Modifying gravity is not an easy task. Any attempt for modifying gravity brings us several theoretical and observational problems which in some cases we can avoid by just constraining the parameters of the theory. The modifications of gravity are usually divided in two branches. The first one is a modification of the energy-momentum tensor by introducing a scalar field, able to reproduce the accelerated expansion of the universe \cite{Tsuji1}. Such models can be also used for explaining the inflationary phase of the universe before the radiation dominated epoch. The second branch of models, correspond to the modifications of the Einstein-Hilbert action by introducing new degrees of freedom. One of the most popular approaches for modifying gravity in this way is the so-called massive gravity theory, which provides a massive term for the graviton \cite{Fierz}. At the linear level, the theory fails due to the additional attractive effect produced by the coupling between the scalar component and the trace of the energy-momentum tensor when the mass of the graviton goes to zero \cite{vDVZ}. The discontinuity disappears at the non-linear level due to the Vainshtein mechanism \cite{Vainshtein}. It was however discovered that non-linearities can introduce a ghost in the theory \cite{Deser}. The first ghost-free formulation of massive gravity at the non-linear level is the so-called dRGT massive gravity \cite{derham}. In this theory, the diffeomorphism invariance is recovered by introducing redundant variables called St\"uckelberg fields, able to restore the gauge invariance of the total action. The theory then requires the introduction of an auxiliary metric called "fiducial" where the extra-degrees of freedom can in principle be stored. It is however, always possible to introduce all the degrees of freedom (5 in total) inside the dynamical metric. In such a case, the fiducial metric is just Minkowski with no-degrees of freedom. The recovery or extension of the diffeomorphism invariance and a ghost-free formulation at the non-linear level, do not guarantee that the theory is free of pathologies. Already some problems have been reported at the cosmological level \cite{DeFelice}. In \cite{Kodama} some apparent problems related to the predictability character of the theory were reported when the linear perturbations around the Schwarzschild de-Sitter solution were analyzed. This apparent pathology is related to the time-direction of the dynamical metric when it contains all the degrees of freedom of the theory and as a consequence, it is related to the definitions of energy inside the dRGT formulation of massive gravity \cite{Miaumiaumiau}. In this manuscript, I demonstrate that at the background level, as far as we have a symmetry under time-translations, it is possible to extend the notion of energy in dRGT and then the equations for a three-dimensional motion will not change with respect to the equations obtained in General Relativity (GR). Then I analyze the Hawking radiation for the generic spherically symmetric black-hole solution obtained in \cite{Kodama} for the case $\beta=(3/4)\alpha^2$, where the cosmological constant ($\Lambda$) is zero but the extra-degrees of freedom are still present. By using the path integral formulation, it is demonstrated that the periodicity associated with the standard time-coordinate is not necessarily related to the periodicity of the "gauge" function $T_0(r,t)\approx t+A(r,t)$, which stores the extra-degrees of freedom inside the spatial-temporal dependence of $A(r,t)$. If the involved periodicities are not the same, then the relation between the rates of emission and absorption will provide an extra-contribution for the black-hole radiation coming from the extra-degrees of freedom of the theory.

\section{The Schwarzschild de-Sitter space in General Relativity}   

The Schwarzschild-de Sitter metric in static coordinates, is defined in agreement with:

\begin{equation}   \label{eq:Sdsm}
ds^2=-e^{\nu(r)}dt^2+e^{-\nu(r)}dr^2+r^2d\theta^2+r^2\sin^2\theta d\phi^2,
\end{equation}

\begin{equation}   \label{eq:e}
e^{\nu(r)}=1-\frac{r_s}{r}-\frac{r^2}{3r_\varLambda^2},
\end{equation}

where $r_s=2GM$ is the gravitational radius and $r_\Lambda=\frac{1}{\sqrt{\Lambda}}$ defines the cosmological constant scale. The equation of motion of a massive test particle in this metric is given by \cite{Mine, Bull1}:

\begin{equation}   \label{eq:Moveq2}
\frac{1}{2}\left(\frac{dr}{d\tau}\right)^2+U_{eff}(r)=\frac{1}{2}\left(E^2+\frac{L^2}{3r_\varLambda^2}-1\right)=C,
\end{equation}

where $C$ is a constant of motion. The effective potential $U_{eff}(r)$, which influences the motion of the test particle, is defined as:

\begin{equation}   \label{eq:effpotaa}
U_{eff}(r)=-\frac{r_s}{2r}-\frac{1}{6}\frac{r^2}{r_\varLambda^2}+\frac{L^2}{2r^2}-\frac{r_sL^2}{2r^3}.
\end{equation}

This potential is clearly independent of the velocity of the test particle. 

\section{The Schwarzschild de-Sitter solution in dRGT} 

Some black-hole solutions corresponding to different metrics have been found in \cite{Gaba}. Additionally, in \cite{Kodama}, the S-dS solution was derived generically for different parameters. All the solutions can be defined as:

\begin{equation}   \label{eq:drgtsolutionbla}
ds^2=g_{tt}dt^2+g_{rr}dr^2+g_{rt}(drdt+dtdr)+r^2d\Omega_2^2,
\end{equation}
 
where:

\begin{eqnarray}   \label{eq:drgt metric}
g_{tt}=-f(r)(\partial_tT_0(r,t))^2,\;\;g_{rr}=-f(r)(\partial_rT_0(r,t))^2+\frac{1}{f(r)},\nonumber\\
g_{tr}=-f(r)\partial_tT_0(r,t)\partial_rT_0(r,t),
\end{eqnarray}

with $f(r)=1-\frac{2GM}{r}-\frac{1}{3}\Lambda r^2$. The metric (\ref{eq:drgt metric}), contains all the degrees of freedom (5 in total) of the theory. In such a case, the fiducial metric is just the Minkowskian one given explicitly as:

\begin{equation}
f_{\mu\nu}dx^\mu dx^\nu=-dt^2+\frac{dr^2}{S_0^2}+\frac{r^2}{S_0^2}(d\theta^2+r^2sin^2\theta),
\end{equation}
 
where $S_0=\frac{\alpha}{\alpha+1}$. The St\"uckelberg fields take the standard form defined in \cite{Kodama}. Note that the solution (\ref{eq:drgtsolutionbla}) can be expressed generically as:

\begin{equation}   \label{eq:Alejandro}
ds^2=-f(r)dT_0^2(r,t)+\frac{1}{f(r)}dr^2+r^2d\Omega^2,
\end{equation}

taking into account that the St\"uckelberg trick allows the extra-degrees of freedom to enter in a similar fashion as the coordinate transformations are performed in GR. At the non-linear level, the extra-degrees of freedom enter as:

\begin{equation}   \label{eq:metric}
g_{\mu\nu}=\left(\frac{\partial Y^\alpha}{\partial x^\mu}\right)\left(\frac{\partial Y^\beta}{\partial x^\nu}\right)g'_{\alpha\beta},
\end{equation}  

with the definitions:

\begin{equation}   \label{eq:Alejandro2}
Y^0(r,t)=T_0(r,t),\;\;\;\;\;Y^r(r,t)=r.
\end{equation} 

Take into account that in general:

\begin{equation}
dt\to\partial_r T_0(r,t)dr+\partial_tT_0(r,t)dt,
\end{equation}

which looks like a coordinate transformation in GR. However, we have to keep in mind that the St\"uckelberg trick is not a diffeomorphism transformation but rather an artifact in order to restore the diffeomorphism invariance of the theory after introducing redundant variables.
    
\section{The effective potential in dRGT massive gravity}   

In order to compare massive gravity with GR, we have to derive the equations of motion for a massive test particle when it moves around a spherically symmetric source. In order to perform the appropriate analysis, it is necessary to work in unitary gauge. Note that the metric (\ref{eq:metric}) with all the degrees of freedom is diffeomorphism invariant under the transformations:

\begin{equation}   \label{eq:gt}
g_{\mu\nu}\to\frac{\partial f^\alpha}{\partial x^\mu}\frac{\partial f^\beta}{\partial x^\nu}g_{\alpha\beta}(f(x)), \;\;\;\;\;Y^\mu(x)\to f^{-1}(Y(x))^\mu.
\end{equation}

The first set of transformations (the left hand-side) corresponds to the usual diffeomorphism transformations in GR. They look similar to the way how the extra-degrees of freedom enter in the theory in agreement with eq. (\ref{eq:metric}). The equations of motion in this case are:

\begin{eqnarray}   \label{eq:Miau}
\frac{1}{2}\left(\frac{dr}{d\tau}\right)^2-\left(\frac{\partial_tT_0(r,t)\partial_rT_0(r,t)}{(\partial_rT_0(r,t))^2-\frac{1}{f(r)^2}}\right)\left(\frac{dr}{d\tau}\right)\frac{E}{g_{tt}}\nonumber\\
+\frac{L^2}{2r^2g_{rr}}=\frac{1}{2g_{rr}}+\frac{E^2}{g_{rr}g_{tt}},
\end{eqnarray}

where $g_{tt}$ and $g_{rr}$ are defined in eq. (\ref{eq:drgt metric}). Note that as $\partial_rT_0(r,t)=0$, the previous equation is reduced to the result (\ref{eq:Moveq2}). If we replace the metric components (\ref{eq:drgt metric}) inside (\ref{eq:Miau}), then we get explicitly:

\begin{eqnarray}   \label{eq:Miau3}
\frac{1}{2}\left(\frac{dr}{d\tau}\right)^2+\frac{\partial_rT_0(r,t)f(r)E}{\partial_tT_0(r,t)(f(r)^2(\partial_rT_0(r,t))^2-1)}\left(\frac{dr}{d\tau}\right)\nonumber\\
-\frac{L^2}{2r^2}\left(\frac{f(r)}{f(r)^2(\partial_rT_0(r,t))^2-1}\right)\nonumber\\
=\frac{f(r)}{2(f(r)^2(\partial_rT_0(r,t))^2-1)}\left(\frac{E^2}{f(r)(\partial_tT_0(r,t))^2}+1\right).
\end{eqnarray}

In eq. (\ref{eq:Miau}), the energy and angular momentum have been introduced in the usual sense in agreement with the results of the first section of the manuscript. The presence of a quantity linear in the velocity of the test particle in eq. (\ref{eq:Miau3}) shows that the effective potential which influences the motion, is velocity-dependent or equivalently, the total energy as it is defined usually is velocity-dependent. This dependence cannot be gauged away as in GR. The origin of the linear velocity term inside the effective potential (or dependence of the total energy with the velocity), comes from the contributions of the extra-degrees of freedom.

\section{Conserved quantities for a test particle moving in dRGT}   

Inside the dRGT formulation of massive gravity, the quantity:

\begin{equation}   \label{eq:Falcao}
g_{\mu \nu}U^\mu U^\nu=C,
\end{equation}

is a constant of motion. It represents the Lagrangian of a test particle moving around a source. If we expand it, then we get the eq. (\ref{eq:Miau}) after taking into account the corresponding conserved quantities. Here I will analyze the constants of motion. By explicit expansion of eq. (\ref{eq:Falcao}), we get:

\begin{eqnarray}   \label{eq:Messi}
g_{tt}\left(\frac{dt}{d\tau}\right)^2+g_{rr}\left(\frac{dr}{d\tau}\right)^2+2g_{tr}\left(\frac{dr}{d\tau}\right)\left(\frac{dt}{d\tau}\right)\nonumber\\
+g_{\phi\phi}\left(\frac{d\phi}{d\tau}\right)^2=C,
\end{eqnarray}

where I have omitted the zenithal angle represented by $\theta$ because we can fix it due to the spherical symmetry of the metric. If we assume the metric to be stationary, then the gauge-transformation function $T_0(r,t)$ is linear in time and then the components of the metric ($g_{\mu\nu}$) are time-independent. In such a case, from eq. (\ref{eq:Messi}), we can find the equations of motion for $t$ and $\phi$ as:

\begin{equation}   \label{eq:CR7}
\frac{d}{d\tau}\left(g_{tt}\left(\frac{dt}{d\tau}\right)+g_{rt}\left(\frac{dr}{d\tau}\right)\right)=0,
\end{equation}

\begin{equation}   \label{eq:CR8}
\frac{d}{d\tau}\left(r^2\left(\frac{d\phi}{d\tau}\right)\right)=0.
\end{equation}

The second equation is just the conservation of the angular momentum. The first one would correspond to the conservation of the total energy as in GR if the term $g_{rt}$ vanishes. From eq. (\ref{eq:CR7}), the total energy is not conserved in its original form, namely, $E=g_{tt}dt/d\tau$. Instead, the conserved quantity is the following combination:

\begin{equation}   \label{eq:Ronaldo}
g_{tt}\left(\frac{dt}{d\tau}\right)+g_{rt}\left(\frac{dr}{d\tau}\right)=E_{dRGT},
\end{equation}

where the subindex dRGT suggests that this quantity should be recognized as an extended total energy inside dRGT. Eq. (\ref{eq:Ronaldo}) however, suggests that the total energy in its usual form is a velocity-dependent quantity. For different values of $dr/d\tau$, the value of $E$ changes. Then any attempt for describing the motion of a particle by using the standard notion of energy is problematic. However, if the conserved energy is extended to the result (\ref{eq:Ronaldo}), it is possible to demonstrate that the equations of motion for a test particle moving around a spherically symmetric source in dRGT, will be identical to the equations of motion obtained from GR. The differences between GR and dRGT will appear when dynamical processes are considered, i.e, processes where the time-coordinate becomes relevant. In the next section I will explore one of such situations.  

\section{The Hawking radiation in dRGT massive gravity}   \label{eq:blabla}

In this section I explore one dynamical process, namely, the Hawking radiation by using the path integral formulation. The St\"uckelberg trick allows us to introduce the extra-degrees of freedom in a similar way as the coordinate transformations are performed in GR in agreement with eq. (\ref{eq:metric}). The "gauge" functions $Y^\alpha(x)$ contain the extra-degrees of freedom information through the spatial and temporal dependence. Here $T_0(r,t)$ plays the role of this function inside the metric (\ref{eq:Alejandro}) as has been demonstrated in eq. (\ref{eq:Alejandro2}). Then at the moment of analyzing dynamical situations, all what we have to do is to replace the standard time-coordinate by the corresponding function $T_0(r,t)$. For example, if we want to define advanced and retarded coordinates in dRGT, we have:

\begin{equation}
V=T_0(r,t)+r+2Mlog\left\vert\frac{r}{2M}-1\right\vert,
\end{equation}

\begin{equation}   \label{eq:extended ret}
U=T_0(r,t)-r-2Mlog\left\vert\frac{r}{2M}-1\right\vert.
\end{equation}   

These coordinate definitions will help us to understand better the Hawking radiation effects in dRGT massive gravity. Here I will focus on the case $\beta=(3/4)\alpha^2$ for the generic solution (\ref{eq:Alejandro}). This case corresponds to a zero cosmological constant ($\Lambda$) but still keeping the contribution coming from the extra-degrees of freedom as has been demonstrated in \cite{Kodama}. 

\section{Analyticity properties of the propagator and the periodicity of the poles}

The differential definition of the propagator is given by:

\begin{equation}   \label{eq:waveequation}
(\square^2-m^2)K(x,x')=-\delta(x,x'),
\end{equation}

with the appropriate boundary conditions. In the Schwarzschild geometry extended to dRGT, we will consider the case where $x'$ is external to the black-hole and $x$ is over the horizon as in the standard case. In dRGT, we can also define the Kruskal coordinates in agreement with:

\begin{equation}   \label{eq:condo1}
ds^2=-\left(\frac{32M^3e^{-r/2GM}}{r}\right)dU'dV'+r^2d\Omega^2,
\end{equation}

\begin{equation}   \label{eq:condo}
U'V'=\left(1-\frac{r}{2GM}\right)e^{r/2GM},
\end{equation}

where the extended version of these coordinates are defined as:

\begin{equation}   \label{eq:adv}
V'=\left(\frac{r}{2GM}-1\right)^{1/2}e^{\left(r+T_0(r,t)\right)/4GM},
\end{equation}

\begin{equation*}   
U'=-\left(\frac{r}{2GM}-1\right)^{1/2}e^{\left(r-T_0(r,t)\right)/4GM}.
\end{equation*}        

Note that in these definitions what really differs with respect to GR, is the inclusion of the function $T_0(r,t)$ instead of the standard time-coordinate $t$. In GR, we trivially have $T_0(r,t)=t$ after gauge transformations. In dRGT, the extra-degrees of freedom enter in the theory in a similar way as the gauge fields enter in GR. Then $T_0(r,t)$ cannot be trivially transformed to the standard time-coordinate. We can observe that the definition of the event horizon does not change with respect to the standard case because the product $U'V'$ is independent of $T_0(r,t)$. We can complexify the event horizon by analytically extending the coordinate $U'$ (or $V'$) to complex values after sending to zero the coordinate $V'$ (or $U'$). Note that the event horizon for an asymptotically flat case ($\beta=(3/4)\alpha^2$ in dRGT) implies $r_{H}=2GM$ and then $U'=0$ or $V'=0$ from eq. (\ref{eq:adv}). The previous coordinates are all defined with respect to $T_0(r,t)$, then the extra-degrees of freedom contribution appears implicitly inside the "gauge" function. From the standard procedures followed in \cite{Hartle}, it is true that any geodesic starting from real values of $x'$, will intercept the horizon at real sections, corresponding to real values of the extended coordinates $U'$ and $V'$. 

\begin{figure}
	\centering
		\includegraphics[width=0.4\textwidth, natwidth=560, natheight=300]{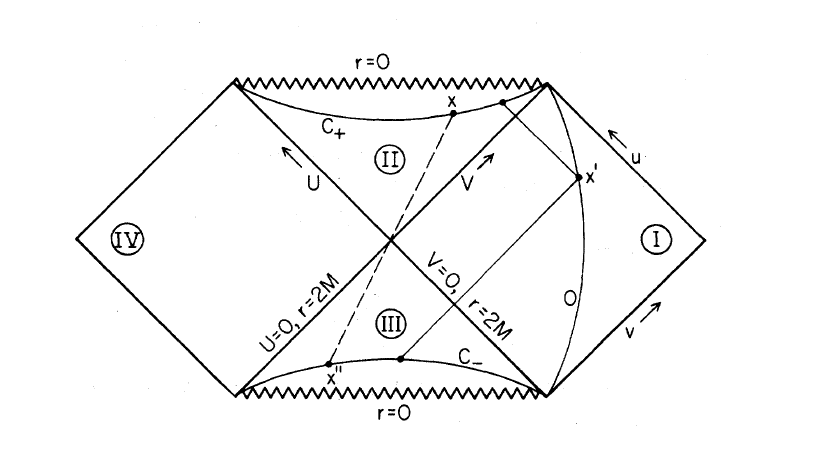}
	\caption{The Penrose diagram for the Schwarzschild geometry. In dRGT, a larger number of paths might appear due to the extra-degrees of freedom. Taken from \cite{Hartle}.}
	\label{fig:momoko}
\end{figure}

If we analyze the propagator properly, we will find that the poles are related to the null geodesics with respect to $T_0(r,t)$ and then the singularities of the propagator are given by $s(x,x')=-i\epsilon$. Here $s(x,x')$ corresponds to null geodesics if we write the metric in terms of the function $T_0(r,t)$, or equivalently, it corresponds to time-like geodesics (almost null) if we write the metric with respect to the usual time coordinate $t$ with the extra-degrees of freedom appearing explicitly through the function $A(r,t)$. In the diagram (\ref{fig:momoko}), the region $II$ contains the Cauchy data in a portion of the future event horizon and another portion of the past event horizon. This region is covered by the coordinates defined in eqns. (\ref{eq:adv}) and the propagator is in this form, defined uniquely inside it. Analogous conclusions can be done for the other regions of the diagram. If we complexify the "gauge" function $T_0(r,t)$ by keeping the ordinary coordinates $r, \theta, \phi$ real, then the coordinates $U'$ and $V'$ can be written as:

\begin{equation}
U'=\vert U'\vert e^{-i\psi(r,t)/4M}, \;\;\;\;\;\;\; V'=\vert V'\vert e^{i\psi(r,t)/4M},
\end{equation}

with $T_0(r,t)=\gamma(r,t)+i\psi(r,t)$. Then the Cauchy data is regular as:

\begin{equation}   \label{eq:mylove}
-4\pi M<\psi(r,t)<0.
\end{equation} 
          				
Then the problem of determining the propagator is reduced to solve the wave equation (\ref{eq:waveequation}) in the real coordinates $\vert U'\vert$ and $\vert V'\vert$ for fixed values of $\psi(r,t)$. The periodicity given by eq. (\ref{eq:mylove}) means that the propagator defined in terms of the extended coordinates $U'(T_0(r,t),r)$ and $V'(T_0(r,t),r)$ will be analytical over the complexified horizon on the upper half-plane with respect to the complex variable $U'$ and it will also be analytical on the lower half-plane with respect to the extended variable $V'$. In eq. (\ref{eq:mylove}), $\psi(r,t)=\mu+\bar{A}(r,t)$, with $\bar{A}(r,t)$ being the complex conjugate of $A(r,t)$. In addition $\mu$ corresponds to the usual imaginary part of the time-coordinate. This means that the analyticity conditions of the propagator will be different if we define it with respect to the time-coordinate $t$ or with respect to the function $T_0(r,t)$. If we define the analyticity conditions with respect to $t$, then the eq. (\ref{eq:mylove}) is equivalent to:

\begin{equation}
-4\pi M-\bar{A}(r,t)<\mu<-\bar{A}(r,t),
\end{equation}			

with $\mu$ being the complex part of the time-coordinate. In the present case, the diagram (\ref{fig:momoko}) represents the vision of an observer defining the events in agreement with the extended coordinates $U'$ and $V'$, namely, written with respect to $T_0(r,t)$. Such observers will not perceive any difference with respect to GR. Physically this is a consequence of the fact that the number of paths involved in the evaluation of the integrals is the same as in GR.
On the other hand, observers defining the time-coordinate in agreement with the usual notion $t$, will perceive contributions coming from additional paths with respect to the GR case. The analytic continuation $T_0(r,t)\to T_0(r,t)-i4\pi M$, relates the point $x$ with the point $x''$ of the diagram (\ref{fig:momoko}). This is equivalent to a relation between the amplitudes for emission and absorption of the black-hole. The observers defining the time in agreement with the function $T_0(r,t)$, will relate the rates of emission and absorption in the usual way as in GR \cite{Hartle}. On the other hand, if the observers define the time-coordinate in agreement with $t$, then this relation is modified and it is equivalent to an extra-component of radiation coming from the extra-degrees of freedom.  

\section{The periodicity of the poles of the propagator}        

\begin{figure}
	\centering
		\includegraphics[width=0.4\textwidth, natwidth=560, natheight=350]{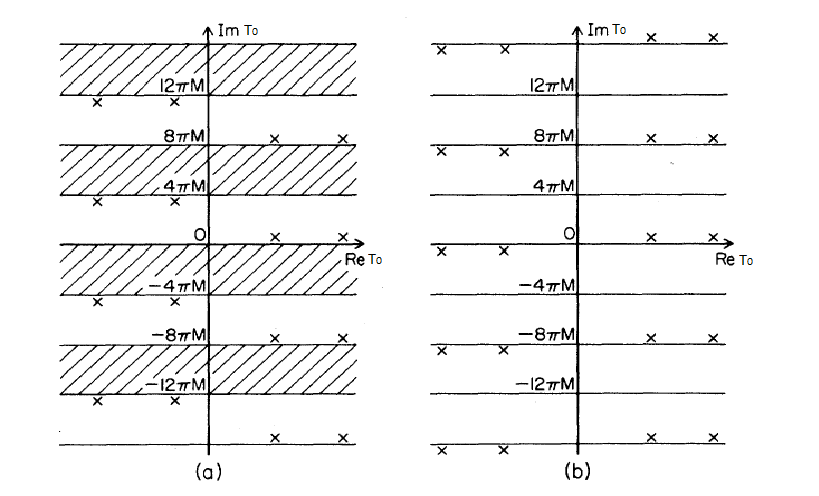}
	\caption{a). The periodicity associated to the propagator for a path starting in region I and then going toward region II. The shaded regions correspond to the analyticity in $T_0(r,t)$. The crosses locate the singularities as the real null geodesics (in the extended sense) connecting a point of the region I, with another one with fixed $r$, $\theta$ and $\phi$ of region II for the case of the singularities located above the real axis. For the same situation, but with the singularities located below the real axis, the connection is between the regions I and III. b). The same analytic structure, but this time for paths starting and finishing in the same region I. Taken from \cite{Hartle}.}
	\label{fig:momoko3}
\end{figure}

\begin{figure}
	\centering
		\includegraphics[width=0.4\textwidth, natwidth=560, natheight=300]{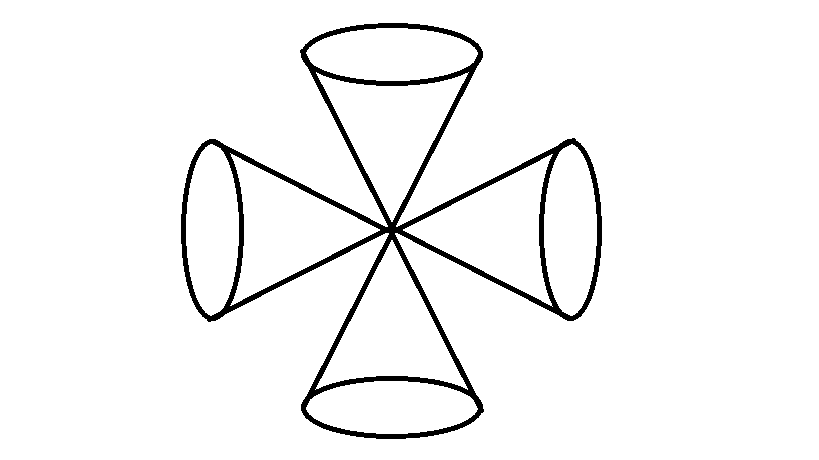}
	\caption{Multiplicity of cones starting at the same spacetime position. Due to the presence of the function $A(r,t)$, what corresponds to a periodicity behavior with respect to the variable $T_0(r,t)$, does not necessarily corresponds to the same periodicity associated to the usual notion of time $t$.}
	\label{fig:momoko2}
\end{figure}

We can easily calculate the positions of the poles of the propagator. In agreement with the previous analysis of the analyticity of the propagator, there are poles when $s(x,x')=-i\epsilon$. It is well known that the poles of the propagator will be located when \cite{Hartle}:

\begin{equation}   \label{eq:Tto}
T_0(r,t)-T_0(r,t)'=\pm\left(\vert\vec{x}-\vec{x}'-i\epsilon\vert\right),
\end{equation}   

or equivalently:

\begin{equation}   \label{eq:Tto2}
t-t'=\pm\left(\vert\vec{x}-\vec{x}'-i\epsilon\vert\right)-(A(r,t)-A'(r,t)),
\end{equation}

is satisfied. The singularities of the propagator in agreement with eq. (\ref{eq:Tto}), become periodic as has been mentioned in the analysis of the previous section. It is the periodicity of the poles what reproduces the effect of particle creation. Since the periodicity associated to the variable $T_0(r,t)$, does not necessarily implies periodicity with respect to $t$ in agreement with the results obtained in eqns. (\ref{eq:Tto}) and (\ref{eq:Tto2}); then an observer describing the physics in agreement with the time $t$ will perceive an extra-component of radiation whenever the extra-degrees of freedom are relevant. This happens after the Vainshtein scale $r_V$. In fact, the periodicity associated to the variable $T_0(r,t)$, has to be divided by two: 1). One fraction belonging to the complex time variable $\mu$. 2). Another fraction going to the complex part of the function $\bar{A}(r,t)$ which is just the complex conjugate function of $A(r,t)$. Then the perception of temperature will depend on whether the detectors are working with $T_0(r,t)$ or $t$ as the notion of time. Two pictures will help us to understand what is going on. The first one is the periodicity associated to the poles of the propagator as they are defined in Fig. (\ref{fig:momoko3}). This case corresponds to the observers using the detectors working with $T_0(r,t)$ as the time-coordinate.
Each cross in the diagram corresponds to the singularities of the propagator and they are related to the null geodesics with respect to the function $T_0(r,t)$. The propagator has a period $8\pi M$ with respect to the function $T_0(r,t)$. This period will not necessarily be the same with respect to the variable $t$. As an example, if we send a ray of light from a point $x'$ outside the horizon toward the future event horizon, its relation with another null ray departing from the same point but going toward the past event horizon is related to the rotation $i4\pi M$ on the complex plane. From the point of view of an observer defining the time in agreement to the variable $t$ and located at scales where the extra-degrees of freedom are relevant, this relation is non-trivial. The observer will perceive just a partial rotation on the complex plane instead of a full rotation in agreement with the periodicity associated to the propagator (periodicity in the sense of $T_0(r,t)$). In fact, depending on the way how the full rotation $i4\pi M$ is distributed between $\mu$ and $\bar{A}(r,t)$, then the observer using $t$ as the standard time, will believe that the ray going to the future event horizon is in reality connected to a multiplicity of geodesics partially rotated with respect to the initial one and connecting different points. The Fig. (\ref{fig:momoko2}) illustrate this effect. The effect of multiplicity of geodesics is equivalent to the change in periodicity associated to the variable $t$. The periodicity might change from point to point in the complex plane for $\mu$, depending on the function $A(r,t)$. This change on the periodicity associated to the propagator, when it is expressed as a function of time $t$, is what create the effect of the extra-particle creation and as a consequence, the changes for the black-hole temperature if it is measured with respect to the detectors operating by using the variable $t$ as the time-coordinate. The periodicity of the propagator with the extra-degrees of freedom contribution appearing explicitly becomes:

\begin{equation}   \label{eq:My lovemomoko}
\mu+\bar{A}(r,t)=4\pi M.
\end{equation}     

If a detector measures the temperature by using as a time-coordinate the "gauge" function $T_0(r,t)$, then the temperature will be $T=(1/8\pi M)$ in agreement with GR. However, if the temperature is perceived by a detector using as a reference the time $t$, then it will be given by: 

\begin{equation}   \label{eq:kappamiau} 
T=\frac{\kappa}{2\pi}=\frac{1}{2\pi(\mu/\pi)}.
\end{equation}

With the relation (\ref{eq:My lovemomoko}), this previous result is equivalent to:

\begin{equation}   \label{eq:temperature}
T=\frac{1}{8\pi M-2\bar{A}(r,t)/\pi},
\end{equation}
  
where $\bar{A}(r,t)$ is the analytically extended component of $A(r,t)$. The corresponding relation between the rates of emission and absorption will be:

\begin{equation}   \label{eq:non-trivial}
N(E)=P(E)e^{-2\pi E(4M-\bar{A}(r,t)/\pi)}=P(E)e^{-2\pi E/\kappa},
\end{equation} 

where  $N(E)$ corresponds to the probability of emission of particles and $P(E)$ corresponds to the probability of absorption. Then what the detector peaked with respect to $t$ perceives, is a contribution coming from the event horizon of the black-hole and an extra-contribution coming from the extra-degrees of freedom. If instead of working under the reflection of the variable $T_0(r,t)$ in order to relate the emission and absorption, we initially perform the reflection with respect to $t$ as $t\to t-i4\pi M$, then analogous results would be obtained but with the roles of $t$ and $T_0(r,t)$ just exchanged.

\section{Gauge-invariant formulation}  
In the previous sections we have analyzed that the observation of an extra-component of radiation coming from the extra-degrees of freedom depends on how the observer defines his/her time in order to perform the measurement. On the other hand, if we go back to the St\"uckelberg language, the preferred direction of time disappears. In such a case, the perception of extra-particles would depend on the relative directions between the ``time''-like Killing vectors defined in the directions $T_0(r,t)$ ($K^{T_0(r,t)}$ for the dynamical metric) and $t$ ($K^t$ for the fiducial metric) respectively in unitary gauge. If both vectors are parallel, then we cannot expect any contribution coming from the extra-degrees of freedom. When the extra-degrees of freedom become relevant, $K^{T_0(r,t)}$ is in general not parallel to $K^t$. Depending on the St\"uckelberg fields configuration, $K^{T_0(r,t)}$ can become null at some scale, after which it becomes space-like if we compare it with the direction of $K^t$. Then in principle, dRGT allows the possibility of getting a Killing horizon at scales where there is no event horizon. In any other gauge, the relative direction between the ``time''-like Killing vectors for both metrics will reproduce the same effects.


\begin{thebibliography}{0}
\bibitem{1}  Carroll S., Living Rev. Rel. 4, (2001) 1.
\bibitem{Tsuji1} Copeland E. J., Liddle A. R. and Wands D. Phys. Rev. D {\bf 57} 4686, (1998); Tsujikawa S., Class. Quantum Grav. 30 214003, (2013).
\bibitem{Fierz} Fierz M. and Pauli W., Proc. Roy. Soc. Lond. , A{\bf173}, 211 (1939).
\bibitem{vDVZ} Van Dam H. and Veltman M. J. G, Nucl. Phys. {\bf B22}, 397.
\bibitem{Vainshtein} Vainshtein A. I., Phys. Lett., {\bf B39}, 393 (1972).
\bibitem{Deser} Boulware D.G. and Deser S., Phys.Rev., D {\bf6}, 3368 (1972).
\bibitem{derham} de Rham C. and Gabadadze G., Phys. Rev. D {\bf82}, 044020 (2010); de Rham C., Gabadadze G. and Tolley A. J., Phys. Rev. Lett. 106, 231101, (2011); Hinterbichler K. , Rev.Mod.Phys. 84 (2012) 671; C. de Rham, Living Rev.Rel. 17 (2014) 7. 
\bibitem{DeFelice} De Felice A., Gumrukcuoglu A. E., Lin C. and Muhkoyama S., Class. Quant. Grav. 30, 184004, (2013).
\bibitem{Kodama} Kodama H. and Arraut I., Prog. Theor. Exp. Phys. 023E02, (2014). 
\bibitem{Miaumiaumiau} Arraut I., Phys. Rev. D, {\bf90} (2014) 124082.
\bibitem{Mine} Arraut I.,  arXiv:1305.0475 [gr-qc]; Arraut I., arXiv:1311.0732.
\bibitem{Bull1} Stuchl$\acute{i}$k Z., Bull. Astron. Inst. Czech. {\bf34}, (1983) 129. 
\bibitem{Gaba} Koyama K., Niz G. and Tasinato G., Phys. Rev. Lett. 107, 131101 (2011); Koyama K., Niz G. and Tasinato G., Phys. Rev. D {\bf84}, 064033 (2011); Sbisa F., Niz G., Koyama K. and Tasinato G., Phys.\ Rev.\ D {\bf 86}, 024033 (2012); Berezhiani L., Chkareuli G., de Rham C., Gabadadze G. and Tolley A. J., Phys.Rev. D {\bf85} (2012) 044024; Nieuwenhuizen T. M., Phys.Rev. D {\bf84} (2011) 024038.
\bibitem{Hartle} Hartle J. B., Hawking S. W., Phys. Rev. D {\bf13}, 2188, (1.976).
\end{thebibliography}
\end{document}